# Amplified Directional Photoluminescence from CIS Quantum Dots and hBN Quantum Emitters using Tunable BIC Metasurfaces

Omar A. M. Abdelraouf

**Affiliations:**
Institute of Materials Research and Engineering, Agency for Science, Technology, and Research (A*STAR), 2 Fusionopolis Way, #08-03, Innovis, Singapore 138634, Singapore

Corresponding author: Omar A.M. Abdelraouf (email: Omar_Abdelrahman@a-star.edu.sg)

## ABSTRACT

Integrated and tunable light sources are critical for advancing quantum nanophotonic chips in quantum computing, communications, and sensing. However, efficient and tunable emission amplification post-fabrication poses major challenges. Hybrid metasurfaces combining niobium pentoxide ($Nb_2O_5$), copper indium sulfide (CIS) quantum dots or hexagonal boron nitride (hBN), and antimony trisulfide ($Sb_2S_3$) as a low-loss phase-change material offer a compelling solution for dynamic control and amplification of photoluminescence and quantum light emission. In this work, we report an active hybrid metasurface supporting tunable bound states in the continuum (BIC) resonances in the visible regime, achieving experimental $Q$-factors up to 206 and strong amplification of CIS QDs photoluminescence as well as quantum light emission of hBN single-photon emitters. The metasurface enables BIC resonance shifts of 33.5 nm in the visible spectrum via phase transition of $Sb_2S_3$, and 17 nm through dimensional parametric tuning. We experimentally demonstrate highly directional photoluminescence amplification up to 33-fold, alongside broad tunable amplified PL emission upon $Sb_2S_3$ phase modulation. Furthermore, we propose amplified, tunable, and on-demand strong coupling of hBN single-photon emitters with the tunable BIC metasurface for next-generation broadband quantum nanophotonic chips. This work sets a new benchmark in reconfigurable nanophotonic platforms for efficient quantum light sources in integrated photonic systems.

**KEYWORDs**: Reconfigurable Nanophotonic Devices, Active Metasurfaces, Bound States in the Continuum, Copper Indium Sulfide Quantum Dots (CIS QDs), Hexagonal Boron Nitride Single-Photon Emitters (hBN SPE), Dielectric Metasurfaces, Phase-Change Materials, Quantum Light Emission, Directional Photoluminescence Amplification

**INTRODUCTION**

Copper indium sulfide quantum dots (CIS QDs) have emerged as promising candidates for high-efficiency photoluminescence applications, demonstrating exceptional external quantum efficiency (EQE) in the visible spectrum with tunable emission wavelengths from 600-920 nm. These non-toxic, environmentally benign quantum dots exhibit superior photoluminescence quantum yields compared to traditional cadmium-based systems while maintaining excellent stability and narrow emission linewidths.[1-3] Concurrently, hexagonal boron nitride (hBN) as a two-dimensional quantum material has garnered significant attention for quantum nanophotonic devices due to its unique quantum light emission properties. The functionalized hBN single photon emitters (SPE) enable full-spectrum emission across the visible range with engineering SPE defects through oxygen plasma treatment.[4-6] This quantum light emission capability, combined with the material's exceptional stability and biocompatibility, positions hBN as an ideal platform for future quantum nanophotonic architectures.[7] However, a fundamental limitation persists in planar structures of light emitting materials results in inherently weak photoluminescence amplification due to limited light-matter interaction and the absence of post-fabrication tunability, which constrains device optimization and practical implementation.

Nanophotonics offers transformative solutions for enhancing light-matter interactions through the precise engineering of resonant optical modes that confine electromagnetic fields to subwavelength volumes of several optoelectronic devices.[8-24] The strategic manipulation of photonic nanostructures enables unprecedented control over spontaneous emission rates, allowing for significant photoluminescence amplification through cavity quantum electrodynamics effects. Among various resonant modes, bound states in the continuum (BIC) have emerged as particularly advantageous for light-emitting devices due to their theoretically infinite quality factors and strong field confinement capabilities.[25-27] BIC resonances provide exceptional enhancement of radiative emission rates while maintaining low optical losses, making them ideal for photoluminescence amplification applications.[28-30] Recent advances have demonstrated BIC-enhanced light-emitting devices achieving remarkable performance improvements, including enhanced emission rates and improved directional control.

The significance of tunable nanophotonic devices extends beyond static performance optimization, offering unprecedented opportunities for simultaneous photoluminescence amplification and post-fabrication tuning capabilities. Traditional nanophotonic structures suffer from fixed resonant frequencies determined during fabrication, limiting their adaptability to different quantum emitters or operating conditions.[31] Post-fabrication tuning enables dynamic optimization of spectral matching between cavity resonances and emitter transitions, maximizing light-matter coupling efficiency. Antimony trisulfide ($Sb_2S_3$) has recently emerged as a revolutionary low-loss phase-change material (PCM) ideally suited for visible wavelength applications. This wide-bandgap semiconductor exhibits extraordinary properties including ultralow optical losses (extinction coefficient $k < 0.01$), large refractive index ($n > 3.5$), and exceptionally gigahertz optical switching speeds.[32-34] The material's transparency window spanning 610 nm to near-infrared, combined with its non-volatile switching characteristics, establishes $Sb_2S_3$ as the optimal choice for tunable nanophotonic applications requiring precise spectral control.

Quantum light emission amplification represents a critical frontier in quantum nanophotonics, where individual photons must be efficiently coupled to optical cavities for enhanced emission rates and improved collection efficiency. Early investigations utilizing nanophotonic cavities for quantum emitter enhancement have demonstrated significant improvements in spontaneous emission rates, with Purcell factors exceeding $10^4$ with strong emission rate amplification.[35] However, a persistent challenge emerges from spectral mismatching between single-photon emission wavelengths and cavity resonances after nanofabrication, which substantially reduces amplification efficiency and limits device performance.[36] This wavelength mismatch problem becomes particularly pronounced in solid-state quantum emitters, where inhomogeneous broadening and spectral diffusion can shift emission frequencies away from optimal cavity resonances. Tunable nanophotonic approaches offer the ideal solution for achieving broadband quantum nanophotonic chips by enabling real-time spectral alignment between quantum emitters and cavity modes, thus ensuring consistent enhancement across diverse quantum systems and facilitating the development of scalable quantum photonic networks.

In this work, we realize an active hybrid metasurface that supports tunable BIC resonances in the visible regime by integrating high-refractive-index $Nb_2O_5$ with CIS QDs for PL amplification or hBN for SPE amplification using low-loss phase-change $Sb_2S_3$.

This metasurface achieves an experimental BIC $Q$-factor of 206, enabling strong and tunable amplification for PL of CIS QDs and quantum light emission enhancement of hBN. We demonstrate that precise tuning of the BIC resonance can be achieved through both phase transitions of $Sb_2S_3$, with resonance shifts up to 33.5 nm, and parametric structural sweeps yielding shifts of 17 nm. The hybrid platform delivers directional PL amplification in CIS QDs up to 33-fold compared to planar films, as well as tunable amplified emission wavelength shifts by 28 nm upon phase change. Furthermore, we propose amplified, tunable, and on-demand coupling of hBN single-photon emitters with the metasurface, establishing a pathway for scalable, broadband quantum nanophotonic chips and integrated reconfigurable photonic systems.[37]

**RESULTS AND DISCUSSION**

Figure 1 presents the conceptual framework of an active metasurface operating in the visible spectrum that supports BIC resonances for compact tunable photoluminescence and quantum light sources. The three-dimensional schematic of the proposed active metasurface is depicted in Fig. 1a. Each unit cell (meta-atom) comprises dual trapezoidal structures fabricated from low-loss $Nb_2O_5$ material. The complete metasurface architecture incorporates a 10 nm $Al_2O_3$ layer and a 130 nm $Sb_2S_3$ layer beneath the dielectric metasurface. CIS quantum dots serve as the gain medium for photoluminescence enhancement and are deposited atop the $Nb_2O_5$ metasurface. Horizontally polarized continuous-wave laser illumination along the x-direction excites the CIS quantum dots, while tunable amplified photoluminescence is collected from the upper surface.

The nanofabrication sequence begins with radio frequency sputtering to deposit 130 nm-thick $Sb_2S_3$ films onto quartz substrates, followed by atomic layer deposition of 10 nm $Al_2O_3$ for phase-change material encapsulation. Electron beam lithography defines the $Nb_2O_5$ metasurface patterns, subsequently processed via ALD deposition and blanket inductively coupled plasma reactive ion etching to remove excess material and residual resist. Comprehensive nanofabrication details are provided in the Methods Section and Figure S2 of the Supporting Information (SI). Figure 1b displays scanning electron microscopy images of the fabricated $Nb_2O_5$ metasurface prior to CIS quantum dot deposition.

Finite-difference time-domain simulations enabled geometric optimization of the hybrid metasurface parameters, including periodicities in $x$ and $y$ directions ($P_x$ and $P_y$), bottom width ($w_2$), height ($H$), and trapezoidal length ($L$), to maximize interference between resonant optical modes within the trapezoidal elements and establish robust BIC resonances. Subsequently, these parameters remained fixed while varying the $y$-direction gap ($g_y$) and top width ($w_1$) of the trapezoids to control light-matter interactions within the active metasurface. Detailed FDTD simulation procedures are outlined in the Methods Section and Figure S1 in the SI.

ymmetry breaking occurs through reduction of the top width ($w_1$) relative to the bottom width by $\Delta w = w_2 - w_1$, as illustrated in Fig. 1c. This asymmetry enables high Q-factor quasi-BIC resonances within the lattice configuration.[25, 38, 39] At $\Delta w = 0$, normal incidence transmission remains high without resonant features due to decoupling of true BIC modes from the radiation continuum, eliminating leaky radiative channels. Upon introducing asymmetry with $\Delta w = 20$ nm in both $Nb_2O_5$ pillars during the amorphous $Sb_2S_3$ state, pronounced quasi-BIC resonances emerge in the radiation continuum. According to the state of PCM material, we named this BIC resonance $BIC_{a-Sb_2S_3}$. This BIC mode, designated according to the phase-change material state, exhibits simulation-predicted formation at 630 nm wavelength with $Q$-factor of 1260.

$Sb_2S_3$ demonstrates substantial refractive index contrast ($\Delta n = 1.2$) between amorphous and crystalline phases.[40] Simulations reveal that transitioning $Sb_2S_3$ from amorphous to crystalline state within the active metasurface induces increased refractive index, causing 30 nm wavelength redshift of $BIC_{c-Sb_2S_3}$ resonance while reducing the $Q$-factor to 228, as shown in Fig. 1c. Furthermore, continuous wavelength tuning of active quasi-BICs in transmission spectra becomes achievable through intermediate $Sb_2S_3$ states.

Experimental optical transmission measurements of the active metasurface are presented in Fig. 1d. During the amorphous $Sb_2S_3$ state, observed BIC resonances closely match simulation predictions from Fig. 1c. Measured BIC resonances exhibit $Q$-factors of 206 at 626.5 nm wavelength. Slight spectral position shifts compared to simulations result from fabrication-induced curved $Nb_2O_5$ pillar edges[41] and interfacial effects between stacked materials.[42] Light scattering losses could be minimized through nanofabrication process optimization. Following thermal phase transition of $Sb_2S_3$ from amorphous to crystalline state using hotplate treatment (300°C, 3 minutes), repeated transmission measurements revealed 33.5 nm redshift in active BIC resonances, consistent with simulation predictions. Measured $Q$-factors of the active BIC resonance decreased to 42, with degradation attributed to increased optical losses in the crystalline phase-change material state.

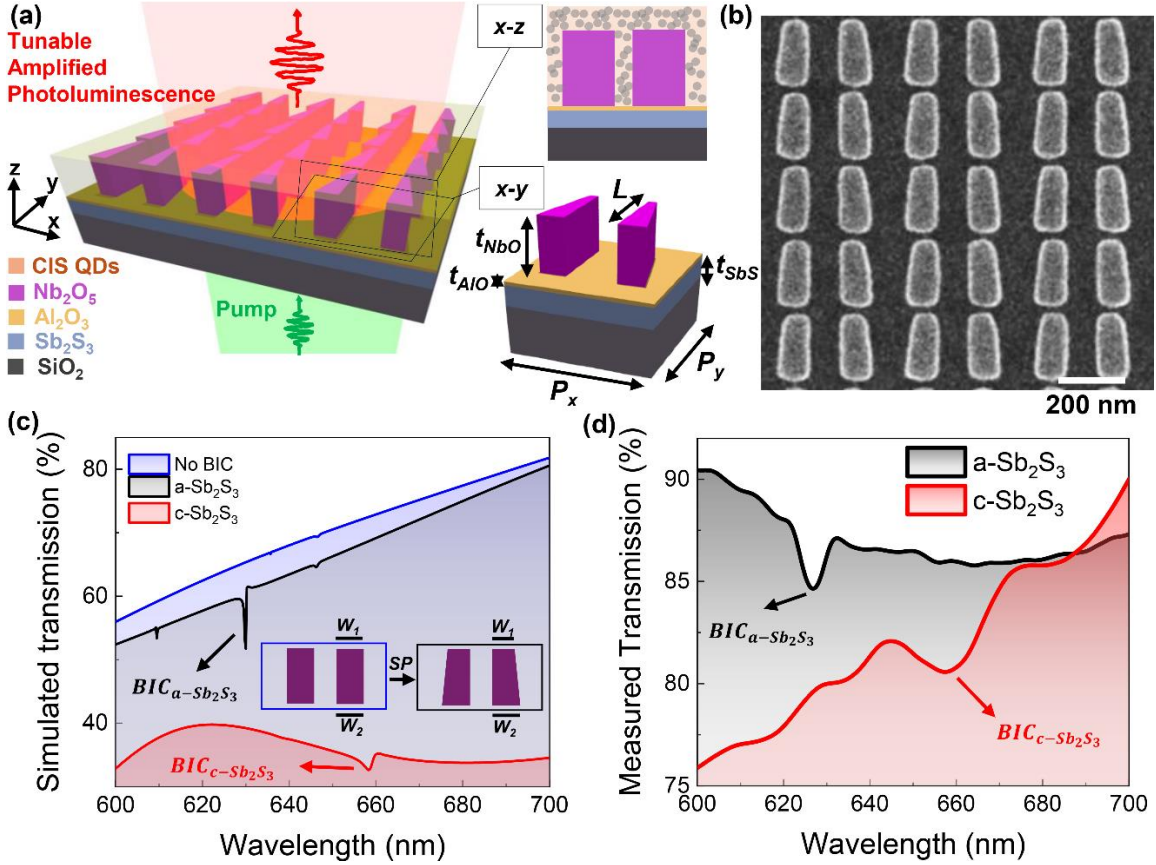

**Figure 1.** Proposed hybrid metasurfaces supporting quasi bound states in the continuum (q-BICs) resonance for tunable amplified photoluminescence of $CuInS_2/ZnS$ quantum dots (CIS QDs) in the visible regime. (a) 3D schematic of the hybrid active metasurface, consisting of $Nb_2O_5$ resonant nanostructures on active material $Sb_2S_3$ using quartz substrate. Optical pumping using green laser (wavelength 532 nm) from the backside illuminate the metasurface and the emitted amplified photoluminescence is collected from the topside. (Inset) 3D meta-atom structure with labelled geometrical parameters including fixed design parameters such as periods in *x*-direction and *y*-direction with values ($P_x$) = 405 nm and *y*-direction ($P_y$) = 234 nm respectively. Thickness of quartz substrate of 500 um, thicknesses of $Sb_2S_3$ ($t_{SbS}$) and $Al_2O_3$ ($t_{AlO}$) films are 130 nm and 10 nm respectively. Height of $Nb_2O_5$ ($t_{NbO}$ = 400 nm). The variable design parameters are length of the trapezoidal shape (*L*), the width of the top and bottom sides ($w_1$ and $w_2$) respectively, and vertical gap between the closed meta-atoms ($g_y = P_y - L$). The *x-z* plane showing the distribution of CIS QDs covering resonant $Nb_2O_5$ nanostructures. (b) Scanning electron microscopy (SEM) images of $Nb_2O_5$ q-BIC cavity without CIS QDs using a scale bar of 200 nm. (c) Simulated transmission without breaking symmetry of hybrid metasurface in blue color. Comparison between simulation transmission after symmetry breaking and changing the phase of $Sb_2S_3$ film from amorphous (a-$Sb_2S_3$) to crystalline (c-$Sb_2S_3$). (d) Measured transmission of hybrid metasurface before (a-$Sb_2S_3$) and after (c-$Sb_2S_3$) phase transition of $Sb_2S_3$.

Multipolar decomposition analysis (MPD) was performed to examine the interference and coupling mechanisms between various optical resonance modes within the hybrid metasurface at BIC wavelengths, as illustrated in Figure 2. During the amorphous Sb₂S₃ phase, MPD calculations presented in Figure 2a demonstrate that active BIC resonance ($BIC_{a-Sb_2S_3}$) originates from coupling between in-plane electric quadrupole (*EQ*) modes and out-of-plane magnetic dipole (*MD*) modes near the BIC wavelength. Comprehensive MPD simulation methodologies are detailed in the Methods Section.

Electric field intensity profiles were computed to analyze field confinement characteristics within Nb₂O₅ structures at BIC wavelengths. Figure 2b displays field intensity distributions across the horizontal (*x-y*) plane at the mid-height of Nb₂O₅ pillars, complemented by vertical (*x-z*) plane distributions spanning the pillar structures. Vector field analysis determined field polarity and directional characteristics within the pillars. At the $BIC_{a-Sb_2S_3}$ wavelength, electric field intensity exhibits strong confinement within the upper Nb₂O₅ pillars primarily through in-plane *EQ* optical modes, while magnetic field confinement occurs via out-of-plane *MD* optical modes surrounding each Nb₂O₅ pillar.

For crystalline Sb₂S₃ conditions, MPD simulations shown in Fig. 2c exhibit redshifted active BIC wavelengths consistent with transmission simulation results. The optical mode strength of the shifted resonance ($BIC_{c-Sb_2S_3}$) diminishes compared to the amorphous state due to increased optical contrast between Nb₂O₅ and crystalline Sb₂S₃, while substrate optical losses reduce field confinement within Nb₂O₅ at these wavelengths. Nevertheless, the tunable resonance characteristics demonstrate potential for achieving continuously adjustable active BIC resonances across broad visible wavelength ranges.[43]

Electric field intensity distributions within Nb₂O₅ at the shifted active BIC wavelength are plotted in Figure 2d. At this wavelength, coupling between in-plane *EQ* and out-of-plane MD modes governs the optical resonance behavior. However, the resulting electric field intensity distribution emerges from combined in-plane *EQ* and *ED* contributions. The enhanced electric field distributions surrounding $Nb_2O_5$ pillars further intensify light-matter interactions, facilitating stronger photoluminescence emission enhancement.

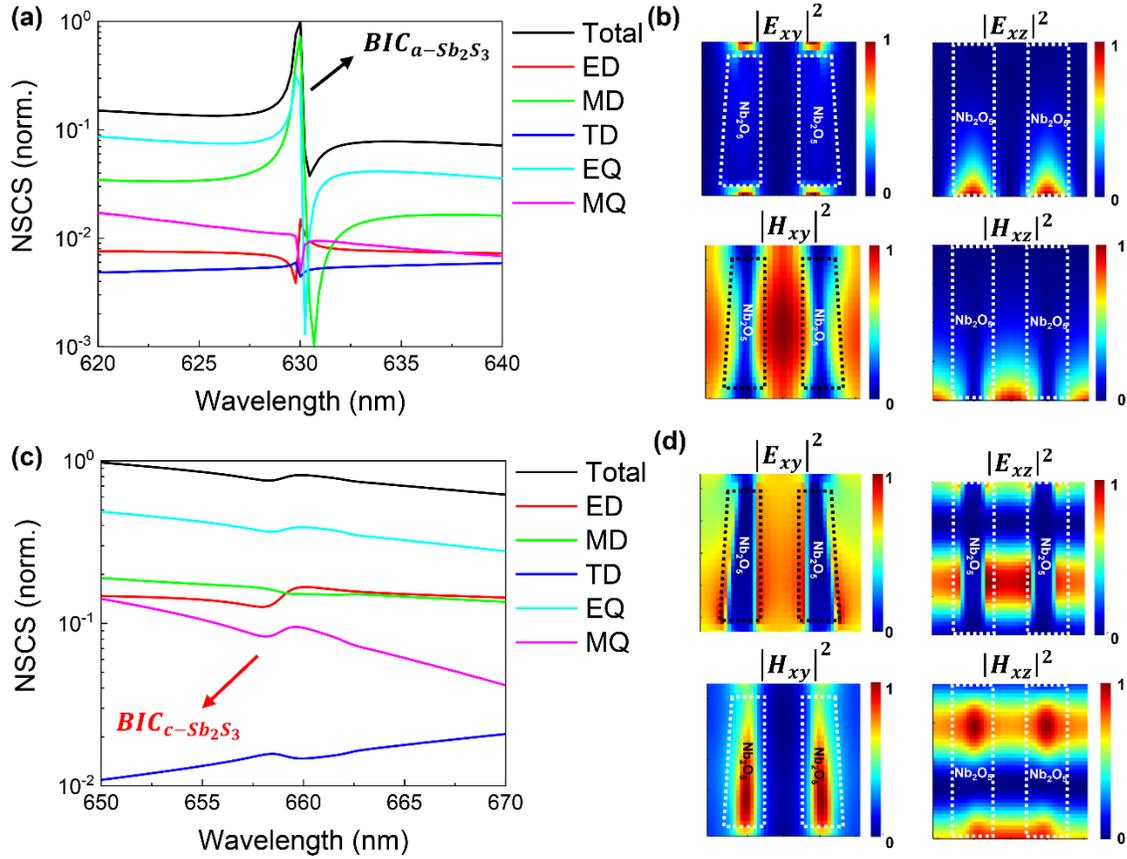

**Figure 2.** Optical multipolar decomposition analysis and electric field confinement simulation of the proposed active hybrid metasurface. (a) Normalized scattering cross section (NSCS) of the hybrid metasurface supporting q-BIC at amorphous state ($BIC_{a-Sb_2S_3}$) using symmetry breaking of $\Delta w$ = 20 nm. Resonant optical modes including the electric dipole (*ED*), the magnetic dipole (*MD*), the toroidal dipole (*TD*), the electric quadrupole (*EQ*), and the magnetic quadrupole (*MQ*). (b) Simulated electric field intensity distribution of amorphous state ($BIC_{a-Sb_2S_3}$) at horizontal plane passing through middle height of $Nb_2O_5$ nanostructure ($|E_{xy}|^2$) and electric field intensity distribution at vertical plane passing through the center of $Nb_2O_5$ nanostructure ($|E_{xz}|^2$). Simulated magnetic field intensity distribution at horizontal plane passing through middle height of $Nb_2O_5$ nanostructure ($|H_{xy}|^2$) and electric field intensity distribution at vertical plane passing through the center of $Nb_2O_5$ nanostructure ($|H_{xz}|^2$). (c) NSCS of the hybrid metasurface supporting q-BIC at crystalline state ($BIC_{c-Sb_2S_3}$) using same symmetry breaking dimension as amorphous. (d) The corresponding simulated electric field intensity distribution and magnetic field intensity distribution of crystalline state ($BIC_{c-Sb_2S_3}$) at the same horizontal and vertical planes.

Beyond modifying $Sb_2S_3$ optical properties for active BIC resonance wavelength tuning, we conducted dimensional parameter sweeps of the gap ($g_y$) and top width ($w_1$) between adjacent meta-atoms during the amorphous $Sb_2S_3$ state, as demonstrated in Fig. 3. Measured transmission spectra for various $g_y$ and $w_1$ configurations are presented in Fig. 3a. Expanding $g_y$ and reducing $w_1$ between meta-atoms decreased both trapezoidal length and top width, producing a blueshift in BIC resonance wavelength from $\lambda$ = 636 nm at $g_y$ = 20 nm and $w_1$ = 70 nm to $\lambda$ = 619 nm at $g_y$ = 30 nm and $w_1$ = 50 nm. Consequently, broadband wavelength tuning capability of up to 17 nm was demonstrated, enabling spectroscopic applications through multiple BIC metasurface arrays with varying dimensional parameters.[44] Corresponding scanning electron microscopy images of the fabricated devices are displayed in Fig. 3b.

Amplified photoluminescence emissions from the fabricated patterns were characterized using a custom optical measurement system. PL enhancement measurement results are shown in Fig. 3c, with the optical setup schematically illustrated in Fig. S4 of the SI. Detailed PL measurement configurations are provided in the Methods Section. Initial characterization focused on PL emission from CIS quantum dot thin films prepared via spin coating on silicon substrates. The measured PL emission spectrum of thin film CIS QDs is shown in Fig. S3.

Measured PL spectra for each fabricated pattern are plotted in Fig. 3c, revealing pronounced PL enhancement at BIC wavelengths that correlates with transmission spectra in Fig. 3a. Normalized PL intensity demonstrates maximum amplification reaching 33-fold compared to planar CIS quantum dot references using $g_y$ = 20 nm and $w_l$ = 70 nm parameters. Alternative dimensional configurations yielded PL enhancement factors ranging from 13 to 24, as measured $Q$-factors vary with different $g_y$ and $w_l$ values.

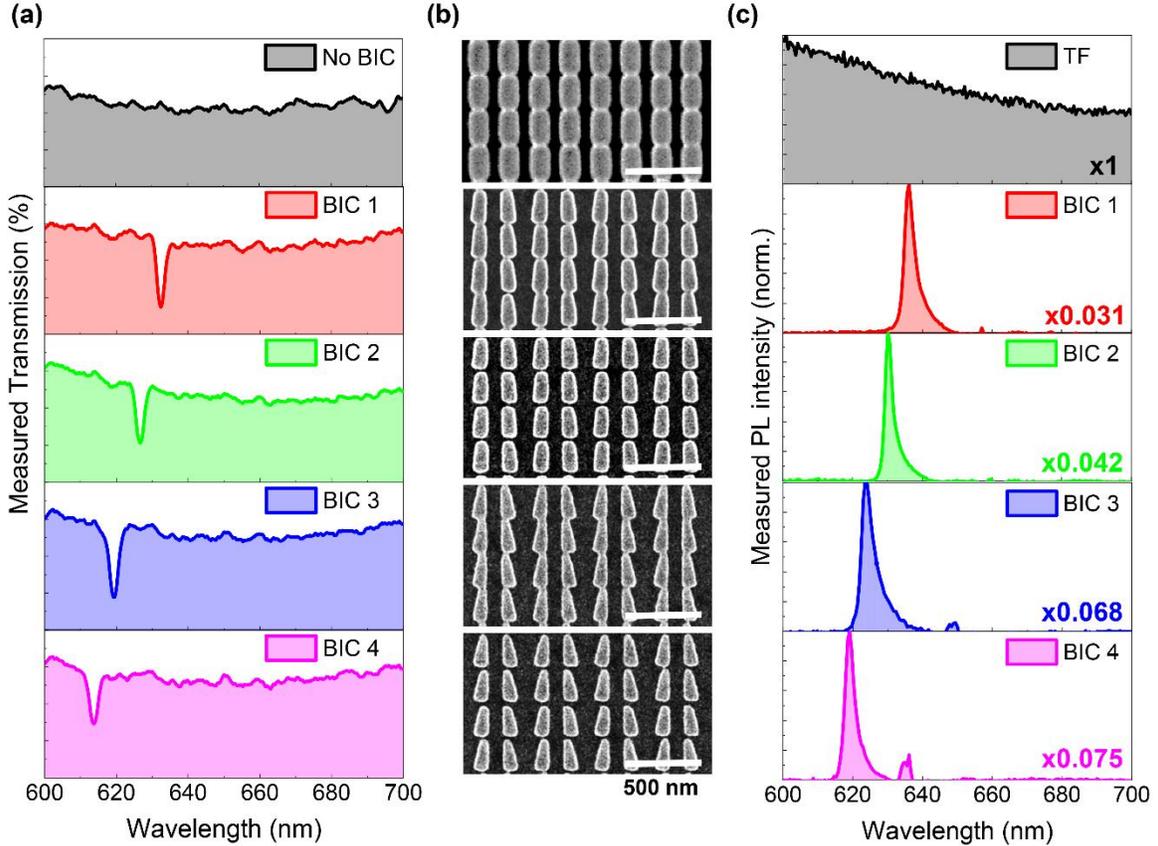

**Figure 3.** Linear optical characterization of the proposed q-BIC hybrid metasurface. (a) Measured transmission of the active hybrid metasurface before symmetry breaking (No BIC) and after symmetry breaking using different vertical gaps ($g_y$) and top width ($w_l$). (b) The corresponding SEM images of the fabricated hybrid metasurface before depositing CIS QDs using scale bar of 500 nm. (c) The measured normalized photoluminescence (PL) spectrum near resonant BIC wavelength range compared with thin film of CIS QDs using same spin coating speed.

Tunable amplified photoluminescence measurements for the highest Q-factor metasurface cavity (specifically, $g_y$ = 20 nm and $w_l$ = 70 nm) are presented in Figure 4. During the amorphous $Sb_2S_3$ phase, the active metasurface exhibits peak emission at 634 nm wavelength with approximately 5 nm full-width half maximum (FWHM). Following thermal phase transformation of $Sb_2S_3$, the tunable amplified photoluminescence peak emission wavelength shifts to 662 nm with ~7 nm FWHM, as illustrated in Fig. 4a. This demonstrates a broad tunable emission wavelength span of 28 nm achieved through $Sb_2S_3$ optical property modulation, confirming the viability of post-fabrication PL tuning and amplification using external stimuli instead of dimensional parameter adjustment.

Furthermore, we characterized amplified PL intensity as a function of pump laser polarization using identical optical configurations. Results for the amorphous state are displayed in Fig. 4b. Pump laser polarization was controlled through half-wave plate rotation. Maximum PL signal occurred under horizontal pump polarization (electric field oriented along the *x*-direction), corresponding to active BIC excitation using *x*-polarized pump illumination as depicted in Fig. 1b. The amplified PL signal disappeared under vertical pump polarization (electric field aligned with the *y*-direction), demonstrating that measured PL exhibits pronounced directionality and enhancement primarily results from strong light-matter coupling between BIC modes and CIS quantum dots surrounding $Nb_2O_5$ pillars. Comparable directional amplified PL behavior was observed during the crystalline phase, as shown in Fig. 4c.

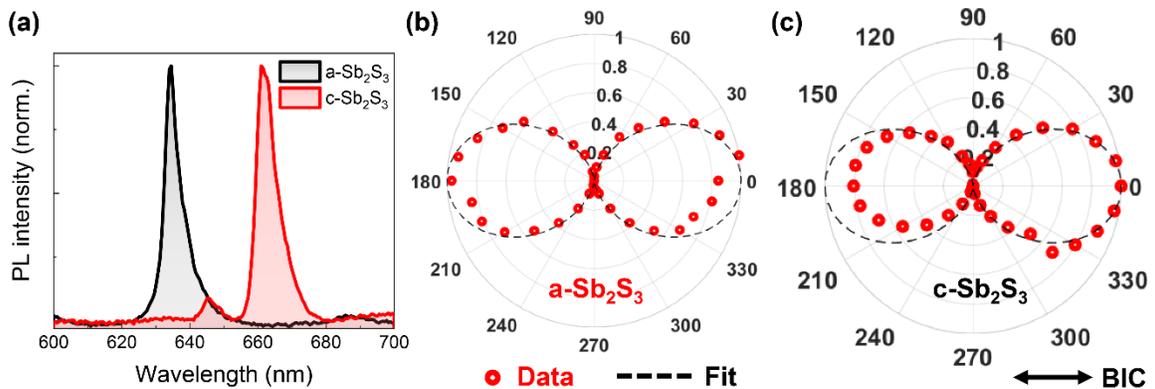

**Figure 4.** Tunable light emission characterization of the proposed q-BIC hybrid metasurface. (a) The normalized amplified PL emission near q-BIC wavelength at amorphous (a-$Sb_2S_3$) and crystalline (c-$Sb_2S_3$) states of the hybrid metasurface coated with CIS QDs. (b) The polar plot of the normalized amplified PL at amorphous state versus different pump polarization angle. (c) The polar plot of the normalized amplified PL at crystalline state versus different pump polarization angle with inset arrow showing the polarization direction of q-BIC in both states.

Single-photon emission in hexagonal boron nitride (hBN) originates from point defects such as carbon substitutions, oxygen plasma treatment, and strain-induced centers within the two-dimensional lattice, each creating a localized electronic transition that emits at a characteristic zero-phonon line.[6, 45, 46] However, the exact emission wavelength of these defect-based SPEs varies randomly across the visible range due to local strain fields, defect type, and dielectric environment. This stochastic distribution makes it impossible to design a passive resonant BIC cavity with a fixed resonance that reliably matches every SPE. Compounding this challenge, individual hBN SPEs exhibit exceptionally few nanometers emission bandwidths which requires a precise spectral overlap to achieve strong light–matter coupling. An active BIC metasurface that can be tuned post-fabrication across the full SPE wavelength distribution therefore enables on-demand alignment with each narrow emitter line, resulting in enhanced coupling and amplified single-photon output. We address these challenges by implementing an active BIC metasurface platform with continuous wavelength tuning to dynamically match and boost the narrow-band SPEs of hBN defects for future quantum nanophotonic chips.

Figure 5 illustrates the design and performance of the tunable metasurface for on-demand amplified quantum light emission via active BIC resonances. In this configuration, a monolayer hBN flake hosting multiple color centers replaces the CIS quantum dot gain medium and is conformally deposited atop the $Nb_2O_5$ pillars. Each defect exhibits a distinct single-photon emission (SPE) spectrum (Fig. 5b). We modeled an active BIC metasurface incorporating a 1.75 nm-thick hBN flake (Fig. 5c). In the amorphous $Sb_2S_3$ state, a pronounced BIC resonance emerges at 630 nm with an ultrahigh $Q$-factor of 2440, attributable to the low optical loss and high refractive index of hBN, which acts as an efficient resonant mirror. Upon switching $Sb_2S_3$ to its crystalline phase, the BIC wavelength redshifts to 657 nm with $Q$-factor of 222, demonstrating broad tunability. Critically, $Sb_2S_3$ supports multiple intermediate refractive index states through partial crystallization, achieved by controlled laser power density modulation.[16, 18] By adjusting the laser exposure, the $Sb_2S_3$ film attains arbitrary refractive indices between its amorphous and crystalline extremes, enabling continuous tuning of the active BIC resonance to precisely coincide with any hBN SPE line.[8] This capability opens pathways for multifunctional, reconfigurable quantum metasurfaces.

We demonstrate over two orders of magnitude amplification for two representative hBN emitters: SPE1, spectrally aligned with the amorphous-state BIC, and SPE2, aligned with the crystalline-state BIC (Fig. 5d). The normalized enhancement yields strong signal-to-background ratios essential for applications in quantum key distribution, on-chip quantum interconnects, and polarization-encoded single-photon sources.[47-49] By suppressing spurious background luminescence and enabling directional SPE emission, active metasurfaces provide a pivotal platform for scalable, reconfigurable quantum nanophotonic devices.

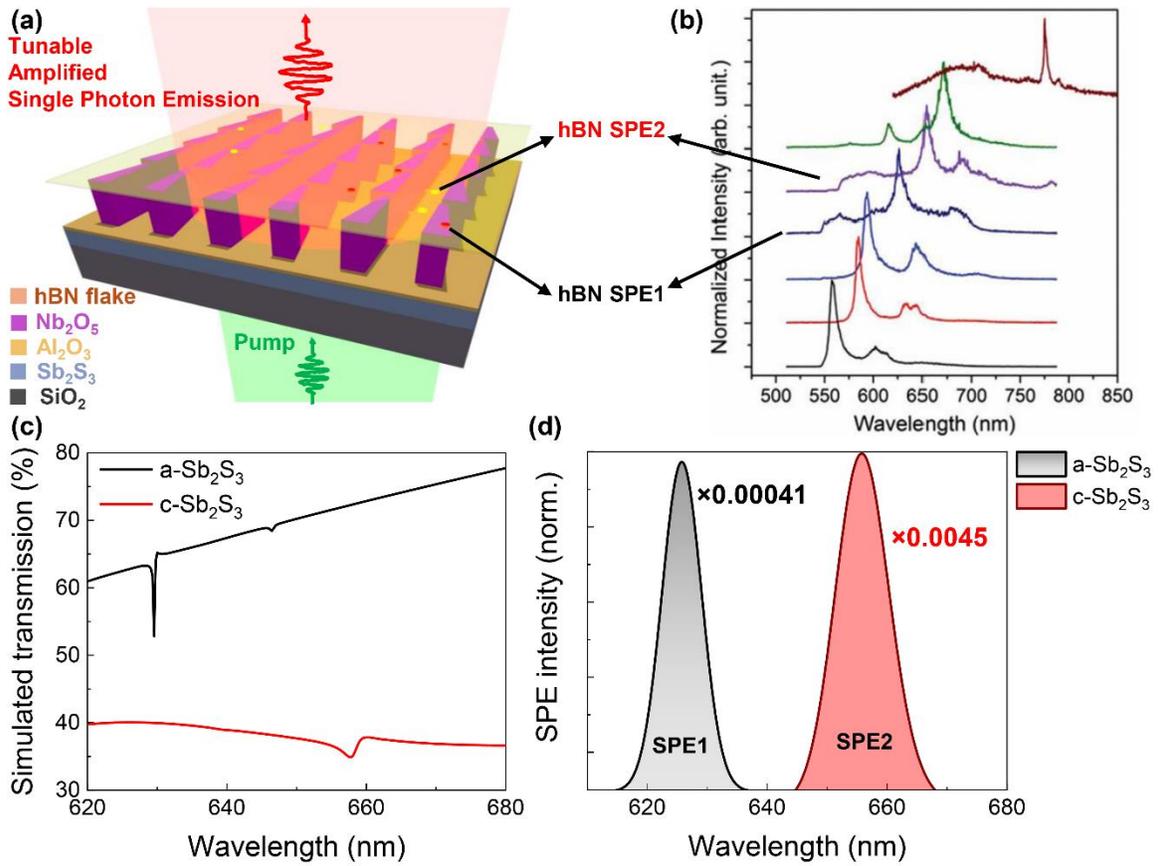

**Figure 5.** The proposed concept for tunable and amplified quantum light emission in 2D materials using hybrid q-BIC metasurface. (a) 3D schematic of the hybrid q-BIC metasurface with staked hexagonal boron nitride (hBN) flake of thickness 1.75 nm with multiple defects as single photon emitters after oxygen ($O_2$) plasma treating. (b) The measured single photon emission spectrum of hBN flake with several defects showing broadband emission wavelength (550 nm to 800 nm). Reproduce with permission from the Royal Society of Chemistry (copyright 2018).[6] (c) The simulated transmission of the hybrid metasurface with stacked hBN flake at amorphous and crystalline states. (d) The simulated amplification of single photon emission of two different emitters at amorphous (SPE 1) and crystalline (SPE 2) states.

## CONCLUSION

In conclusion, we have demonstrated a novel hybrid metasurface platform based on $Nb_2O_5$ integrating CIS QDs or hBN single-photon emitters, and $Sb_2S_3$ phase-change material to achieve tunable BIC resonances with high *Q*-factors directional photoluminescence or SPE amplification. Our experimental results reveal BIC resonances with *Q*-factors up to 206, enabling directional photoluminescence amplification of CIS QDs by a factor of 33 compared to planar films. The metasurface supports dynamic wavelength tuning of the BIC resonance up to 33.5 nm via phase change in $Sb_2S_3$ and 17 nm through parametric modulation, accompanied by a tunable amplified photoluminescence shift of 28 nm. This hybrid metasurface-based approach paves the way for reconfigurable nonlinear and quantum nanophotonic devices.[50-52] Moreover, it underscores the potential in advancing quantum technologies, optical communications, and sophisticated sensing applications, providing a scalable solution for tunable and amplified light emission in integrated photonic devices.[53-55] Future research may explore further device integration using AI-assisted design, wavelength range extension, and exploitation of the strong light-matter interactions for quantum computing and secure communication platforms, accelerating the realization of practical quantum photonic technologies.[56-60]

## METHODS

**Transmission and Electric Field Distribution Profile Simulations**

Optical numerical simulations of the hybrid metasurface were performed using a 3D finite-difference time-domain (FDTD) solver (Lumerical FDTD Solutions).[61] Refractive indices of constituent materials, determined from ellipsometry post-deposition (Figure S1 in SI), were used as input parameters. A linearly polarized plane wave (*x*-axis) illuminated the metasurface structure, with periodic boundary conditions applied in-plane (*x–y* axes) and perfectly matched layers (PML) employed vertically (*z*-axis) to absorb outgoing waves and eliminate artificial reflections. To ensure accurate spatial resolution, the simulation mesh size was set to 10 nm or finer. Field monitors were placed underneath the quartz substrate to record transmission, at the vertical midpoint of the $Nb_2O_5$ trapezoidal pillars to capture horizontal field profiles at BIC resonances, and a vertical monitor at the metasurface center height for detailed field mapping. An auto-shutoff threshold of $5\times10^{-7}$ guaranteed simulation convergence and high precision of results.

**Optical multipolar decomposition calculations**

Multipolar decomposition analysis was performed using the same 3D FDTD simulation environment. Two three-dimensional field monitors were placed around the hybrid metasurface meta-atom cavity. The first monitor captured the electric field vectors throughout the nanoscale metasurface elements at each mesh point, providing spatially resolved electromagnetic data. The second monitor evaluated the cavity's effective refractive index distribution over the entire volume. Using simulated data, we computed the scattering cross-sections corresponding to the electric dipole ($C_{ED}$), electric quadrupole ($C_{EQ}$), magnetic dipole ($C_{MD}$), and magnetic quadrupole ($C_{MQ}$) modes using equations:

$$C_{ED} = \frac{k_0^4}{6\pi\epsilon_0^2 E_0^2} \left| p_{car} + \frac{ik_0}{c}\left(t + \frac{k_0^2}{10}\overline{R_t^2}\right)\right|^2 \quad (1),$$

$$C_{EQ} = \frac{k_0^6}{80\pi\epsilon_0^2 E_0^2} \left| \overline{\overline{Q_e}} + \frac{ik_0}{c}\overline{\overline{Q_t}}\right|^2 \quad (2),$$

$$C_{MD} = \frac{\eta_0^2 k_0^4}{6\pi E_0^2} \left| m_{car} - k_0^2 \overline{R_m^2}\right|^2 \quad (3),$$

$$C_{MQ} = \frac{\eta_0^2 k_0^6}{80\pi E_0^2} \left| \overline{\overline{Q_m}}\right|^2 \quad (4),$$

These scattering components were obtained through well-established multipolar expansion formulas describing the contributions of electric ($p_{car}$), toroidal ($t$), and magnetic dipole ($m_{car}$) moments, as well as electric and magnetic quadrupole moments ($\overline{\overline{Q_e}}$, and $\overline{\overline{Q_m}}$). Detailed derivations for these multipolar components follow the approaches documented in relevant literature.[62]

**Fabrication of hybrid BICs metasurface**

Figure S2 shows the fabrication flow for tunable enhanced light emission in the visible spectrum. Double-sided polished transparent deep-UV quartz substrates (Photonik Singapore) were cleaned by sequential rinsing with acetone, isopropanol (IPA), and acetone. The substrates underwent additional cleaning in an ultrasonic bath for 10 minutes while immersed in acetone, followed by brief rinses with acetone and IPA. Antimony trisulfide ($Sb_2S_3$) films of 130 nm thickness were deposited by RF sputtering (UBM) at ambient temperature. Argon plasma at 21.6 sccm flow, 20 W RF power, and 10 mTorr chamber pressure was employed. A protection layer $Al_2O_3$ of thickness 10 nm using atomic layer deposition (ALD, Beneq TFS 200). Using trimethylaluminium (TMA) and water ($H_2O$) precursors, reactions were run for 100 cycles at a chamber pressure of 4 mTorr and a substrate temperature of 80 °C. $N_2$ was used to purge remaining gas precursors in between cycles. ALD deposition done at substrate temperature of 80 °C. Electron beam lithography (EBL) using an Elionix ELS-7000 system created nanoscale molds for niobium pentoxide ($Nb_2O_5$) pillars. Positive resist ZEP520A was spin-coated at 3000 rpm for 90 seconds, achieving 400 nm thickness, then baked at 180°C for 2 minutes. Ezspacer was applied at 1500 rpm for 30 seconds to mitigate charging effects, with excess removed using nitrogen gas. EBL exposure utilized 200 pA beam current at 100 kV acceleration, with 320 μC/cm² base dose over a 300×300 μm² field containing 60,000 exposure points. Post-exposure development began with Ezspacer removal using deionized water and nitrogen drying. The sample was developed in xylene for 30 seconds, then quenched with IPA. $Nb_2O_5$ deposition employed atomic layer deposition (ALD) in a Beneq TFS 200 system, achieving 100 nm thickness using tert-butylimino tris-diethylamino-niobium precursor and $H_2O$ at 85°C substrate temperature (below ZEP520A glass transition at 105°C). The ALD growth rate was approximately 0.59 Å/cycle. Blanket inductively coupled plasma reactive ion etching (ICP-RIE) using an Oxford OIPT Plasmalab system removed the top 100 nm $Nb_2O_5$ layer. Fluoroform ($CHF_3$) plasma at 25 sccm flow, 150 W RF power, 900 W ICP power, and 25 mTorr pressure etched at 150 nm/min rate. Residual ZEP520A resist was removed via oxygen plasma descum using 50 sccm $O_2$ flow at 250 W RF power for 3 minutes.

**Transmission measurement**

Optical transmission characterization was performed using a supercontinuum nanosecond pulsed laser source (Opera, Leukos Laser Inc.) with 30 kHz repetition rate. Linearly polarized illumination along the x-axis was achieved through a polarization control system comprising a quarter-wave plate (QWP), half-wave plate (HWP), and linear polarizer (LP) to excite the dual BIC metasurface resonances. The transmitted optical signal was captured and fiber-coupled for spectral analysis using an Ocean Optics USB4000 spectrometer interfaced with a computer-based data acquisition system.

**PL and polarization measurement**

The photoluminescence (PL) measurement configuration for resonant metasurface characterization is illustrated in Figure S4. A fiber-coupled 532 nm continuous-wave laser (WiTEC Alpha300) with up to 50 mW output power served as the excitation source. Polarization-dependent PL analysis was enabled through a rotatable linear polarizer controlling the pump beam polarization state. The excitation beam was directed via a dichroic mirror to a 5× objective lens (NA = 0.15). Sample positioning and focusing were controlled using a motorized three-axis translation stage enabling two-dimensional PL mapping and axial adjustment. PL collection utilized the same 5× objective in a confocal arrangement. A 532 nm notch filter eliminated residual pump light to protect the detection system. Sample alignment employed white light illumination routed through a beam splitter (BS), with a flip mirror (M2) directing reflected illumination through a focusing lens (L1) to a CCD camera for real-time imaging. Spectral analysis was performed using a WiTEC PL detection system equipped with either 150 or 600 grooves/mm diffraction gratings. Polarization-resolved PL measurements incorporated a linear polarizer positioned before the detector to analyze emission enhancement as a function of polarization angle.

## ASSOCIATED CONTENT

**Supporting Information**

The measured ellipsometer refractive indices of $Nb_2O_5$, $Sb_2S_3$, and $Al_2O_3$; nanofabrication flow for the hybrid metasurface cavity; PL of CIS QDs before and after annealing; Confocal PL measurement setup; Benchmark comparison table with literature works for PL and single photon emission amplification and tunability using nanophotonic cavities.

## CORRESPONDING AUTHORS

Email: Omar_Abdelrahman@a-star.edu.sg

## ORCID

Omar A. M. Abdelraouf: https://orcid.org/0000-0002-9065-7414

## AUTHOR CONTRIBUTIONS

O.A.M.A. conceived the idea, designed the active hybrid BIC metasurface cavity, carried out 3D FDTD simulations and multipolar decomposition, performed cleanroom materials deposition and characterization, nanofabrication of the hybrid BIC metasurface cavity, linear optical measurements, amplified photoluminescence measurements, optical tunning phase change materials, and wrote the manuscript.

**Competing Financial Interests**

The authors declare no competing financial interest.

**ACKNOWLEDGMENT**

Authors acknowledge funding from A*STAR under its Career Development Fund (CDF) grant no. C233312016 and SERC Central Research Funds (CRF). Also, the support from Singapore international graduate award (SINGA).